\documentstyle[aps,prd,epsf]{revtex}
\draft
\begin{document}
\thispagestyle{empty}
%======================================%
%<<<<<<<<<<<< TITLE PAGE >>>>>>>>>>>>>>%
%======================================%
\thispagestyle{empty}
{\baselineskip0pt
\leftline{}
\rightline{\large\today}
}
\vskip20mm
\begin{center}
{\Large\bf Vacuum Brane and the Bulk Dynamics in Dilatonic Brane World}
\end{center}

\begin{center}
{\large Hirotaka Ochiai${}^{1}$ and Katsuhiko Sato${}^{1,2}$ }
\vskip 3mm
\sl{${}^1$Department of Physics, The University of Tokyo, Tokyo 113-0033, Japan
\vskip 5mm
${}^2$Research Centre for the Early Universe (RESCEU), \\ 
The University of Tokyo, Tokyo 113-0033, Japan
}
\end{center}

%\begin{center}
%{\it }
%\end{center}

%======================================%
%<<<<<<<<<<<<< ABSTRACT >>>>>>>>>>>>>>>% 
%======================================%
\begin{abstract} 
We investigate the dynamics of vacuum brane and the bulk
in dilatonic brane world.
We present exact dynamical solutions
which describe the vacuum dilatonic brane world.
We find that the solution has initial singularity and singularity
at spatial infinity.
\end{abstract}
%======================================%
%<<<<<<<<<<<< SECTION I  >>>>>>>>>>>>>>%
%======================================%
\baselineskip25pt
\section{Introduction}
\label{sec:introduction}
In modern theories of unified physical interactions,  
spacetime has more than four dimensions. 
It is well-known that Kaluza-Klein theory~\cite{kaluza,klein}
tells us that the spacetime 
which has more than four dimensions.
Superstring theories (e.g. Ref.~\cite{polchinski} )
are at the moment the most promising candidates
for a unified description of the basic physical interactions.
There are five anomaly-free, perturbative superstring theories.
The critical dimensions of spacetime are ten for these theories.
There is now evidence that the five superstring theories are related 
with each other by dualities.
It is conjectured that these theories can be regarded as
the limits of an unique theory, 
so called M-theory, 
in which spacetime has eleven dimensions.
M-theory seems to be related to $N=1$, $D=11$ supergravity theory
at low energy.

Conventionally, it is thought that 
the extra dimension is quite small, say the Planck scale,
and then it has not been observed yet.

How does compactification of the extra dimension take place?
The explanation is through multidimensional cosmology
~\cite{cd,freund,os}.
In general relativity, the geometry of spacetime is dynamical.
The three-dimensional space we observe was once as small 
as the internal space, and expanded during evolution of the universe,
while internal space contracted or has remained small 
during evolution of the universe.
Therefore, internal space is microscopic and is not observable.
This explanation is called dynamical compactification.

One of alternative approaches 
by branes in string theory
has been suggested~\cite{ah,rs1,rs2}.
That is
the brane world scenario,
where the Standard Model gauge and matter fields exist inside branes 
while gravitons propagate in the bulk of spacetime.
If this is true, the situation 
in which gravity exists in the bulk of the 11-dimensional spacetime,
while the Standard Model particles exist on a 3-brane is possible. 
Phenomenologically, the brane world scenario was proposed
to solve the hierarchy problem
(e. g. Refs.~\cite{ah,rs1}).
This type of the brane world scenario
may be motivated by Ho\v{r}ava and Witten~\cite{hw,witten}.
Ho\v{r}ava-Witten scenario is reduced to five-dimensional effective
theory at low energy.
Indeed,
Lukas and et al~\cite{lukas} have shown that this five-dimensional theory
admits a supersymmetric solution
describing a pair of thin domain walls.
One of the four-dimensional domain walls in five-dimensional spacetime
is considered as our universe.

Randall and Sundrum (RS1)~\cite{rs1} proposed a toy model
to solve the hierarchy problem.
This model is the system with two branes
like the Ho\v{r}ava-Witten scenario.
In addition, Randall and Sundrum (RS2)~\cite{rs2} presented the 
scenario with noncompact extra space.
This model is the system with a single positive tension brane.
In this scenario, although the hierarchy problem is not solved,
it is very interesting
that gravity is localized on 3-brane
due to the exponentially decreasing warp factor of the bulk metric. 
In the basis of the scenario, the brane world cosmology 
has been studied so far~\cite{kl,s,ssm,b,k,i,mukoh,bow,m,kis,ks}.

The low-energy effective theories of superstring theories and M-theory
are supergravity theories,
including dilaton field and antisymmetric tensor field
etc~\cite{polchinski}.
The dilaton field exists naturally in string theories.
Since we might be able to assume that
the antisymmetric field behaves as a cosmological constant effectively,
it is interesting
to investigate dilatonic brane world scenario with a cosmological constant.
The related works have been done so far
~\cite{youm,NO3,lid,ajs,NO2,adks,kss,neu,nick,low,mw,mb,bv,ken}.

We investigate the dilatonic brane world
in this paper.
For simplicity, we concentrate on vacuum brane cases.
The static solutions in this model have been already
investigated in Ref.~\cite{youm}.
On the ground of the static solutions,
we find the dynamical solutions
which may be able to describe an era of the early universe.
The related studies can be seen in~\cite{cve}.

The plan of the paper is as follows.
In $\S$II, we introduce the model of dilatonic brane world.
In $\S$III, we briefly review the static solutions of the model.
In $\S$IV, we find the dynamical solutions of the model. 
In the last section, we give the conclusions and remarks.
%======================================%
%<<<<<<<<<<<< SECTION II >>>>>>>>>>>>>>%
%======================================%
\baselineskip25pt
\section{The model}
We consider a brane world scenario in dilatonic gravity
with a cosmological constant
and interpret
the $(D-1)$-dimensional thin domain wall of the $D$-dimensional spacetime
as a $(D-2)$-brane world.
For simplicity, we consider the cases that the brane is vacuum.
The action of the model in the Einstein frame is given by
\begin{eqnarray}
S=S_{bulk}+S_{brane},\\
S_{bulk}&=&\frac{1}{2}\int d^{D}x \sqrt{-g}
\biggl[R-\frac{4}{D-2}(\partial \phi)^2
-\Lambda e^{\lambda\phi/(D-2)}\biggr],\\
S_{brane}&=&-\frac{1}{2}\int _{{\mbox{\rm brane}}}d^{D-1}x
\sqrt{-q}Ve^{\lambda\phi/2(D-2)}
+S_{YGH},
\end{eqnarray}
where $q$ is the determinant of the induced metric of the brane,
$\lambda$ is the dilaton coupling, 
$\Lambda$ is the cosmological constant in the bulk,
$V$ is the cosmological constant in the brane.
The last term $S_{YGH}$ 
is the York-Gibbons-Hawking boundary term~\cite{york,gh}.
From now on, we assume that a brane is located at $y=0$
and that
the metric is the form
\begin{equation}
ds^2=g_{\alpha\beta}dx^{\alpha}dx^{\beta}
=q_{\mu\nu}dx^{\mu}dx^{\nu}+b(x,y)^2 dy^2,
\end{equation}
where $\alpha$, $\beta$=0, 1, 2, $\cdots$,
$(D-1)$ and $\mu$, $\nu$=0, 1, 2, $\cdots$, $(D-2)$.
The variation of the metric leads us the $D$-dimensional Einstein equations
\begin{equation}
R_{\alpha\beta}-\frac{1}{2}g_{\alpha\beta}R=T_{\alpha\beta}
-\frac{V}{2b}e^{\lambda\phi/2(D-2)}q_{\alpha\beta}\delta(y),
\end{equation}
where the energy-momentum tensor $T_{\alpha\beta}$ in the bulk is given by
\begin{equation}
T_{\alpha\beta}=
\frac{4}{D-2}\biggl[-\frac{1}{2}(\partial\phi)^2g_{\alpha\beta}
+\partial_{\alpha}\phi\partial_{\beta}\phi\biggr]
-\frac{1}{2}\Lambda g_{\alpha\beta}e^{\lambda\phi/(D-2)}.
\end{equation}
The $D$-dimensional dilaton field equation
is also obtained as
\begin{equation}
8\Box \phi
-\lambda \Lambda e^{\lambda\phi/(D-2)}
-\frac{\lambda V}{2b}e^{\lambda\phi/2}\delta(y)
=0.
\end{equation}

%======================================%
%<<<<<<<<<<<< SECTION III >>>>>>>>>>>>>%
%======================================%
\baselineskip25pt
\section{Static solutions}
\label{sec:Static solutions}
Before presenting our dynamical solutions,
we briefly review the static solution~\cite{youm}.
We assume a metric of the form
\begin{equation}
ds^2=a(y)^2\eta_{\mu\nu}dx^{\mu}dx^{\nu}+b(y)^2 dy^2,
\end{equation}
where $\mu$, $\nu$=0, 1, 2, $\cdots$, $(D-2)$ and $\eta_{\mu\nu}$ denotes 
the metric of the ($D-1$)-dimensional Minkowski space.
The variables $a$ and $b$ are
the scale factors of $(D-1)$-dimensional spacetime
and the orbifold, respectively.
The dilatonic field $\phi$ is a function of the coordinate $y$
of the extra space i. e. 
\begin{equation}
\phi=\phi (y).
\end{equation}
The $D$-dimensional Einstein equations are given by
\begin{eqnarray}
(D-2)\frac{a''}{a}+
\frac{(D-2)(D-3)}{2}(\frac{a'}{a})^2
-(D-2)\frac{a'b'}{ab}\\\nonumber
+\frac{2}{(D-2)}(\phi ')^2
+\frac{1}{2}\Lambda b^2 e^{\lambda \phi/(D-2)} 
+\frac{V}{2b}e^{\lambda\phi/2(D-2)}\delta(y)
=0,\\
\frac{1}{2}(D-1)(D-2)(\frac{a'}{a})^2-\frac{2}{D-2}(\phi ')^2
+\frac{1}{2}\Lambda b^2 e^{\lambda \phi/(D-2)}
=0,
\end{eqnarray}
where primes denote the derivatives
with respect to the coordinate $y$ of the orbifold.
The $D$-dimensional dilatonic field equation is described as
\begin{equation}
\phi ''+(D-1)\frac{a'}{a}\phi '-\frac{b'}{b}\phi '
-\frac{\lambda\Lambda}{8}b^2 e^{\lambda \phi/(D-2)}
\\\nonumber
-\frac{\lambda V}{2b}e^{\lambda\phi/2(D-2)}\delta(y)
=0.
\end{equation}
When the orbifold has the $S^1/Z_2$ symmetry,
the junction conditions on the brane ~\cite{is} are given by
\begin{eqnarray}
a'(+0)=\frac{V}{4(D-2)},\\
\phi'(+0)=\frac{\lambda V}{16}.
\end{eqnarray}
The solution satisfying the junction conditions has the following form:
\begin{eqnarray}
a(y)=H^{\frac{2}{(D-2)\Delta}},\\
b(y)=H^{\frac{2(D-1)}{(D-2)\Delta}},\\
\Phi(y)\equiv e^{-\frac{\lambda\phi}{2(D-2)}}
=H^{\frac{\lambda^2}{8(D-2)\Delta}},
\end{eqnarray}
where
\begin{eqnarray}
H=1+Q|y|,\\
\Delta=\frac{\lambda^2-16(D-1)}{8(D-2)}.
\end{eqnarray}
The parameter $Q$ is related to the bulk cosmological constant $\Lambda$
is as
\begin{equation}
\Lambda=\frac{2Q^2}{\Delta}.
\end{equation}
From the junction conditions on the brane,
we obtain the relation between the bulk cosmological constant $\Lambda$
and the brane tension $V$
\begin{equation}
\Lambda=2^{-5}\Delta V^2.
\end{equation}

When $\lambda=0$, this solution corresponds to the Randall-Sundrum 
non-dilatonic solution
and when $\lambda=4\sqrt{6}$, this solution does to
the Ho\v{r}ava-Witten scenario~\cite{hw}.
As shown in Ref.~\cite{youm},
when $\Delta$ is lower than $-2$, 
the graviton is trapped within the brane
and then the Newton gravity is reproduced.
This is expected to the extension of the Randall-Sundrum scenario
~\cite{rs2}.

In the dilatonic case ($\lambda\neq 0$)
with positive $Q$,
the dilaton field becomes singular at $|y|\rightarrow \infty$.
When the parameter $Q$ is negative,
the singularity of the dilaton field appears at $|y|=|Q|^{-1}$.
We prefer $Q>0$ because we are interested in the noncompact extra dimension
like the RS2.
Then, when $-2(D-1)/(D-2)<\Delta<0$,
the spacetime has the curvature singularity at $|y|\rightarrow \infty$. 
By solving the null geodesics equations,
we find that the $|y| \rightarrow\infty$ singularity
occurs at the spatial infinity.
Otherwise, when $\Delta>0$, 
the curvature singularity does not appear
at the singular point of the dilatonic field ($|y|\rightarrow \infty$).
%======================================%
%<<<<<<<<<<<< SECTION IV >>>>>>>>>>>>>>%
%======================================%
\baselineskip25pt
\section{Dynamical solutions}
\label{sec:Dynamical solutions}
In this section we obtain the dynamical solutions
by following the procedure in ~\cite{lukas}.
We assume that the metric was the form :
\begin{equation}
ds^2=
-N(\tau,y)^2 d\tau^2+a(\tau,y)^2 \delta_{mn}dx^{m}dx^{n}+b(\tau,y)^2 dy^2,
\end{equation}
where $m$, $n$=1, 2, $\cdots$, $(D-2)$.
The dilaton field is a function of time coordinate $\tau$ and
the coordinate $y$ of internal space, i. e.
\begin{equation}
\phi=\phi(\tau,y).
\end{equation}
We assume that all of dynamical variables can be separable as
\begin{eqnarray}
N(\tau,y)&=&n(\tau)a(y),\\
a(\tau,y)&=&\alpha(\tau)a(y),\\
b(\tau,y)&=&\beta(\tau)b(y),\\
\phi(\tau,y)&=&\phi_1(\tau)+\phi(y),
\end{eqnarray}
In the above, $a(y)$, $b(y)$ and $\phi(y)$ are the static solutions
in the previous section.
As a result, the Einstein equations are given by
\begin{eqnarray}
\frac{a^2}{b^2}
\biggl[-(D-2)\frac{a''}{a}
-\frac{(D-2)(D-3)}{2}(\frac{a'}{a})^2
+(D-2)\frac{a'b'}{ab}
\\\nonumber
-\frac{2}{D-2}(\phi ')^2
-\frac{1}{2}\Lambda e^{\lambda \phi/(D-2)}\beta^2b^2e^{\lambda \phi_1/(D-2)}
\\\nonumber
-\frac{V}{2b}e^{\lambda \phi/2(D-2)}\beta e^{\lambda\phi_1/2(D-2)}\delta(y)
\biggr]
\\\nonumber
+\frac{\beta ^2}{n^2}
\biggl[\frac{(D-2)(D-3)}{2}\biggl(\frac{\dot{\alpha}}{\alpha}\biggr)^2
+(D-2)\frac{\dot{\alpha}\dot{\beta}}{\alpha\beta}
-\frac{2}{D-2}\dot{\phi}_1^2\biggr]&=&0,
\\
\frac{a^2}{b^2}\biggl[(D-2)\frac{a''}{a}+
\frac{(D-2)(D-3)}{2}(\frac{a'}{a})^2
-(D-2)\frac{a'b'}{ab}
\\\nonumber
+\frac{2}{D-2}(\phi ')^2
+\frac{1}{2}\Lambda e^{\lambda \phi/(D-2)}
\beta^2b^2e^{\lambda \phi_1/(D-2)}
\\\nonumber
+\frac{V}{2b}e^{\lambda \phi/2(D-2)}\beta e^{\lambda\phi_1/2(D-2)}\delta(y)
\biggr]
\\\nonumber
+
\frac{\beta^2}{n^2}
\biggl[-(D-3)\frac{\ddot{\alpha}}{\alpha}-\frac{\ddot{\beta}}{\beta}
-\frac{(D-3)(D-4)}{2}\biggl(\frac{\dot{\alpha}}{\alpha}\biggr)^2
-(D-3)\frac{\dot{\alpha}
\dot{\beta}}{\alpha\beta}
\\\nonumber
+(D-3)\frac{\dot{n}\dot{\alpha}}{n\alpha}
+\frac{\dot{n}\dot{\beta}}{n\beta}
-\frac{2}{D-2}(\dot{\phi_1})^2
\biggr]&=&0,
\\
\frac{a^2}{b^2}\biggl[\frac{1}{2}(D-1)(D-2)\biggl(\frac{a'}{a}\biggr)^2
-\frac{2}{D-2}(\phi')^2
\\\nonumber
+\frac{1}{2}\Lambda e^{\lambda\phi/(D-2)}
b^2\beta^2 e^{\lambda\phi_1/(D-2)}\biggr]
\\\nonumber
+
\frac{\beta^2}{n^2}\biggl[
-(D-2)\frac{\ddot{\alpha}}{\alpha}
-\frac{(D-2)(D-3)}{2}\biggl(\frac{\dot{\alpha}}{\alpha}\biggr)^2
+(D-2)\frac{\dot{n}\dot{\alpha}}{n\alpha}
\\\nonumber
-\frac{2}{D-2}\dot{\phi_1}^2
\biggr]&=&0,\\
(D-2)\frac{a'}{a}\frac{\dot{\beta}}{\beta}-\frac{4}{D-2}\dot{\phi}\phi '=0.
\end{eqnarray}
The $D$-dimensional dilatonic field equation is
\begin{eqnarray}
\frac{a^2}{b^2}\biggl[\phi''+(D-1)\frac{a'}{a}\phi'-
\frac{b'}{b}\phi'
-\frac{\lambda\Lambda}{8} e^{\lambda\phi/(D-2)} e^{\lambda\phi_1/(D-2)}
\\\nonumber
-\frac{\lambda V}{2b}
e^{\lambda\phi/2(D-2)} e^{\lambda\phi_1/2(D-2)}\delta(y)
\biggr]\\\nonumber
-\frac{\beta^2}{n^2}
\biggl[\ddot{\phi_1}+(D-2)\frac{\dot{\alpha}}{\alpha}\dot{\phi_1}
+\frac{\dot{\beta}}{\beta}\dot{\phi_1}-\dot{n}\dot{\phi_1}\biggr]
&=&0.
\end{eqnarray}

From Eq. (31),
we obtain $\beta \propto e^{-\frac{\lambda\phi_1}{2(D-2)}}$.
and set
\begin{equation}
\beta e^{\frac{\lambda\phi_1}{2(D-2)}}=1,
\end{equation}
and then
the equations of motion (Eq. (28-32)) are separated by variables.
The first [  ] of the left hand sides of the Einstein equations
(28)-(30)
and the dilaton field equation (32)
vanishes by using the $D$-dimensional Einstein equations
and the dilaton field equation in static case.

From now on, we solve the time-dependent parts of equations of motion.
We choose the gauge condition as
\begin{equation}
n(\tau)=const.
\end{equation}
The solution is given by
\begin{eqnarray}
\alpha=\alpha_0 |\tau-\tau_{0}|^p,\\\nonumber
\beta=\beta_0 |\tau -\tau_{0}|^q,\\\nonumber
\Phi\equiv e^{-\frac{\lambda\phi_1}{2(D-2)}}
=\beta=\beta_0 |\tau -\tau_{0}|^q,
\end{eqnarray}
where
\begin{eqnarray}
p&=&p_{\pm}\\\nonumber
:&=&\frac{1+(D-2)(4/\lambda)^2\pm\sqrt{1+(D-3)(4/\lambda)^2}}
{D-1+(4(D-2)/\lambda)^2},\\
q&=&q_{\pm}:=-(D-2)p_{\pm}+1.
\end{eqnarray}
The parameter dependence of the power indices $p_{\pm}$, $q_{\pm}$
are shown in the Figs. 1 and 2. 
When $\tau -\tau_{0}$ is negative,
the first solution (($+$) solution)
is that the worldvolume contracts and 
the orbifold expands. 
The second solution (($-$) solution)
is that both the worldvolume and the 
orbifold contracts.
When $\tau -\tau_{0}$ is positive,
the first solution (($+$) solution)
is that the worldvolume expands and 
the orbifold contracts. 
The second solution (($-$) solution)
is that both the worldvolume and the 
orbifold expand.
Since the powers $p$ or $q$ of the scale factors
is smaller than one,
the expansion is subluminal.
This solution may be interpreted
as Friedmann-Robertson-Walker universe.
As shown in the Fig. 1,
as the absolute value of the dilaton coupling parameter is getting larger,
the power indices $p_+$, $q_+$ of the scale factors is getting larger
in the ($+$) solutions.
The dilaton coupling makes the absolute values of Hubble parameters large.
Otherwise, in the ($-$) solutions,
as the absolute value of the dilaton coupling parameter is getting larger,
the index $p_-$ of the scale factor of the external space is getting larger
and that the index $q_-$ of the internal space is getting smaller.

\begin{figure}
%   \psfrag{l}{$\lambda}
%   \includegraphics{dil1.eps}
  \epsfxsize=95mm
  \centerline{\epsfbox{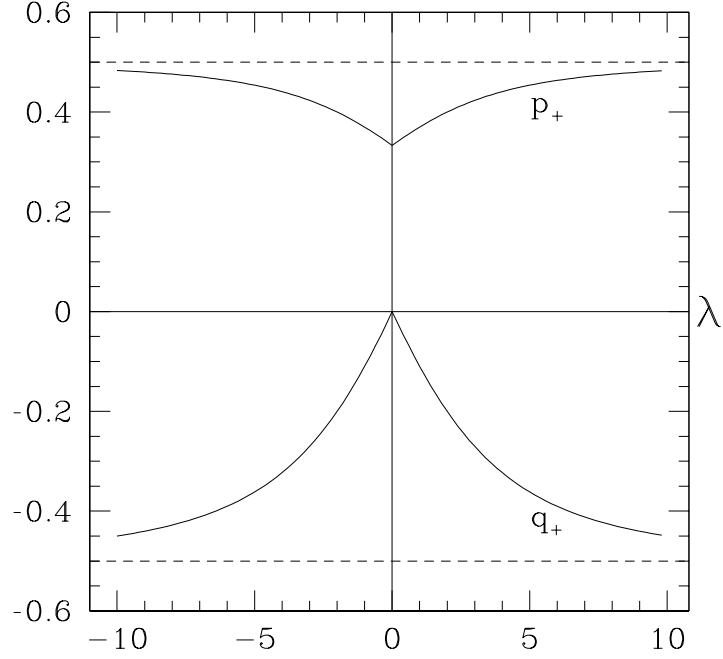}}
\caption{The indices $p_+$, $q_+$ of the dynamical solutions for $D=5$. 
In the non-dilatonic case ($\lambda=0$),
the values of the indices $p_+$, $q_+$ are 1/3 and 0, respectively.
In the Ho\v{r}ava-Witten model ($\lambda=4\sqrt{6}$),
the values of the indices $p_+$, $q_+$ are 
$\frac{9+4\sqrt{3}}{33}$ and $-\frac{4\sqrt{3}-2}{11}$,
respectively.
The values of the indices $p_+$, $q_+$ are asymptotic to $1/2$,
and $-1/2$, respectively,
as the absolute value
of the dilatonic coupling constant is getting larger.
}
\label{fig:4}
\end{figure}

\begin{figure}
  \epsfxsize=95mm
  \centerline{\epsfbox{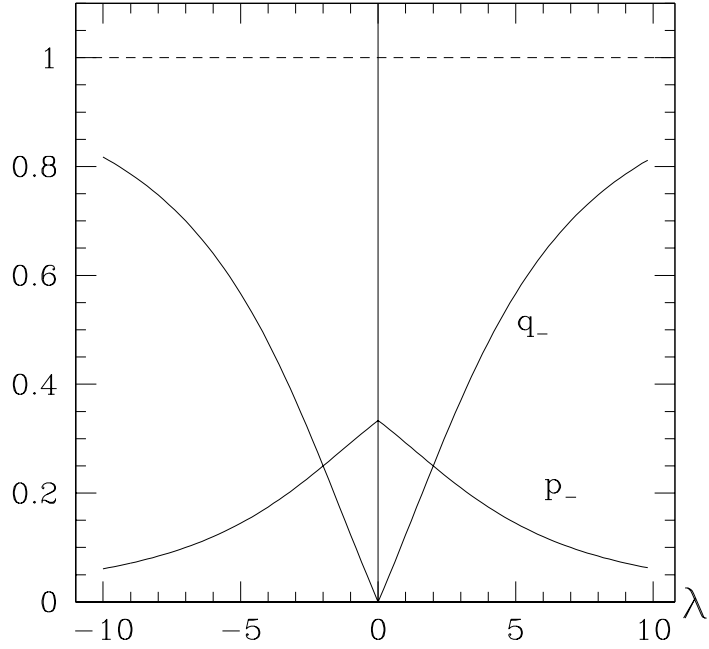}}
\caption{The indices $p_-$, $q_-$ of the dynamical solutions for $D=5$.
In the non-dilatonic case ($\lambda=0$),
the values of the index $p_-$, $q_-$
equal the values of the indices $p_+$, $q_+$
and are $1/3$ and $0$, respectively.
In the Ho\v{r}ava-Witten model ($\lambda=4\sqrt{6}$),
the values of the indices $p_-$, $q_-$ are 
$\frac{9-4\sqrt{3}}{33}$ and $\frac{4\sqrt{3}+2}{11}$,
respectively.
The values of the indices $p_+$, $q_+$ are asymptotic to $0$,
and $1$, respectively,
as the absolute value
of the dilatonic coupling constant is getting larger.
}
\label{fig:5}
\end{figure}

It is illustrative to figure out the Friedmann equation on the brane.
The equation is derived
from the $(0-0)$ component of the $D$-dimensional Einstein equations
(Eq. (28):
\begin{equation}
\frac{(D-2)(D-3)}{2}\biggl(\frac{\dot{\alpha}}{\alpha}\biggr)^2
+(D-2)\frac{\dot{\alpha}\dot{\beta}}{\alpha\beta}
-\frac{2}{D-2}\dot{\phi}_1^2=0.
\label{hubble}
\end{equation}

In Appendix, we give the sketch of the derivation of
the $(D-1)$-dimensional effective gravitational equations
on the brane~\cite{mw,mb,bv}
to obtain to physical meaning of each terms
in the left-hand side of Eq.~(\ref{hubble}).
The effective gravitational equation
is given by:
\begin{eqnarray}
{}^{(D-1)}G_{\mu\nu}
= {4(D-3) \over (D-2)^2}\hat{T}_{\mu\nu}
-{}^{(D-1)}\Lambda q_{\mu\nu} - E_{\mu\nu}, 
\label{4dEinstein2}
\end{eqnarray}
where
\begin{eqnarray}
\hat{T}_{\mu\nu}
&=&D_{\mu}\phi D_{\nu}\phi-\frac{D}{2(D-1)}(D_{\mu}\phi)^2 q_{\mu\nu},
\\
{}^{(D-1)}\Lambda &=& \frac{1}{2}[\Lambda
-\frac{\Delta}{2^5}V^2]e^{\lambda\phi/(D-2)}=0.
\end{eqnarray}
Note that the $(D-1)$-dimensional effective cosmological constant vanihshes
in this model.
In the $(D-1)$-dimensional effective gravitational equations
on the brane,
the term $E_{\mu\nu}$ is the contribution of the bulk to the brane.
In appendix, we derive the expression $E_{\mu\nu}$
in the case with the radion $b$.
For our solutions we evaluate each terms as follows:
\begin{eqnarray}
{}^{(D-1)}G_{00}&=&\frac{(D-2)(D-3)}{2}
\biggl(\frac{\dot{\alpha}}{\alpha}\biggr)^2,\\
\frac{4(D-3)}{D-2}\hat{T}_{\mu\nu}&=&
\frac{2(D-3)}{(D-1)(D-2)}(\dot{\phi_1})^2,\\
%{D-3 \over D-2}\left(T_{\rho\sigma}
%q^{~\rho}_{\mu}  q^{~\sigma}_{\nu}  
%+\left(T_{\rho\sigma}n^\rho n^\sigma-{1 \over D-1}T^\rho_{~\rho}\right)
% q_{\mu\nu} \right) 
%\\\nonumber
%=diag\biggl(\frac{4(D-3)}{(D-2)^2}(\dot{\phi_1})^2, 0, ..., 0\biggr)
%+\frac{D-3}{2(D-1)}
%&\biggl[&\frac{4D}{(D-2)^2}\biggl(\frac{\dot{\phi_1}}{a}\biggr)^2
%+\frac{4}{D-2}\biggl(\frac{\phi'}{\beta b}\biggr)^2
%-\Lambda \beta^{-2}e^{\lambda\phi_0/(D-2)}\biggr]q_{\mu\nu},\\
% KK_{\mu\nu} 
%-K^{~\sigma}_{\mu}K_{\nu\sigma} -{1 \over 2}q_{\mu\nu}
%  \left(K^2-K^{\alpha\beta}K_{\alpha\beta}\right)
%&=&-\frac{(D-2)(D-3)}{2\beta^2 b^2}\biggl(\frac{a'}{a}\biggr)^2 q_{\mu\nu},\\
E_{00}&=&\frac{D-3}{D-1}
\biggl[\frac{\ddot{\alpha}}{\alpha}-\frac{\ddot{\beta}}{\beta}
-\biggl(\frac{\dot{\alpha}}{\alpha}\biggr)^2+
\frac{\dot{\alpha}\dot{\beta}}{\alpha\beta}\biggr]\\
&=&\frac{(D-2)(D-3)}{D-1}\biggl[-\biggl(\frac{\dot{\alpha}}{\alpha}\biggr)^2
+\frac{\dot{\alpha}\dot{\beta}}{\alpha\beta}\biggr].
\end{eqnarray}
%By the static Einstein equation (Eq. (11)),
%the terms which is proportional to the $y$-dependent terms
%are canceled out. 
Using the spatial components of the
$D$-dimensional Einstein equations (Eqs. (29-30)),
we can eliminate the term $\ddot{\alpha}-\ddot{\beta}$ in Eq. (44)
and obtain Eq. (45).
The first term on the left-hand side of Eq. ~(\ref{hubble}) is obtained from
the $(D-1)$-dimensional Einstein tensor ${}^{(D-1)}G_{00}$ and $E_{00}$.
The second term on the left-hand side is obtained from $E_{00}$.
The third term on the left-hand side
corresponds to the $(D-1)$-dimensional energy-momentum tensor
of the dilaton field.
In the work~\cite{ssm},
it is shown that $E_{\mu\nu}$ fall off
and cannot carry away the energy momentum from a system to infinity
in the perturbation theory around the Randal-Sundrum solution.
In our results,
$E_{00}$ affect on the Hubble parameter 
$\frac{\dot{\alpha}}{\alpha}$
and is not negligible.
This is because 
the energy-flow of the dilatonic field from the brane to the bulk
has an effect on $E_{\mu\nu}$.

Finally, we comment the global structure of the bulk geometry.
From the trace of the $D$-dimensional Einstein equation,
the scalar curvature in the bulk is given by
\begin{equation}
R=\frac{4}{D-2}(\nabla \phi)^2+\frac{D}{D-2}\Lambda e^{\lambda\phi/(D-2)}.
\end{equation}
Substituting our solutions to the above,
we obtain
\begin{eqnarray}
R=-\frac{16(D-2)q^2}{\lambda^2}H^{-\frac{4}{(D-2)\Delta}}|\tau -\tau_{0}|^{-2}
\\\nonumber
+\frac{1}{(D-2)(\beta_0)^2}\biggl(D+
\frac{\lambda^2}{8\Delta}\biggr) \Lambda H^{-\frac{\lambda^2}{4(D-2)\Delta}}
|\tau -\tau_{0}|^{-2q}.
\end{eqnarray}
Note that the $|y| \rightarrow \infty$ singularity appears.
As discussed in $\S$III,
the $|y| \rightarrow \infty$ singularity is the curvature singularity
when $-2(D-1)/(D-2)<\Delta<0$,
and is not so when $\Delta>0$.
In addition, the initial singularity exisits at $\tau=\tau_0$,
the big crunch singularity also appears at $|\tau|\rightarrow \infty$
when $q$ is negative.
%======================================%
%<<<<<<<<<<<< SECTION V  >>>>>>>>>>>>>>%
%======================================%
\baselineskip25pt
\section{Discussion and summary}
%%%%%%%%%%%%%%%%%%%%%%%%%%%%%%%%%
\label{sec:summary and discussion}
In this paper we have investigated 
the bulk dynamics
in a dilatonic vacuum brane world with a cosmological constant.
We presented the dynamical bulk-solutions in $\S$IV.
The behavior of the solutions depend
on the dilaton coupling constant $\lambda$.
It is the next issue to investigate the dilatonic brane world
with ordinary matter.

The Randall-Sundrum toy model with noncompact extra space is
given by the limit when the dilaton coupling parameter
vanishes.
In the Randall-Sundrum toy model, 
the orbifold is stabilized i. e. $q=0$.
This scenario can be considered as the valid model for
the universe after the stabilization of the orbifold.
As shown in the case of the static solution,
the nearby solutions of the Randall-Sundrum solution
are extended to the Randall-Sundrum scenario.

Even if the solutions do not correspond to the extended 
Randall-Sundrum scenario,
when the fifth dimension is compact,
the solutions are not necessarily
excluded physically.
When $\tau -\tau_{0}$ is positive,
there are the solutions in which
the four dimensional worldvolume expands
and internal space contracts.
If the fifth dimension is compact,
this solution is approximately reduced to the normal Kaluza-Klein scenario
with the dynamical compactification.
This solution may be a candidate
for a expanding phase of the early universe.

For the explanation of homogeneity and density perturbation,
it is believed that the early universe has been in the inflationary paradime.
Though the dynamical solutions which we have presented in this paper
correspond not to the inflation but to the subluminal expansion,
the inflationary solutions can exist in more realistic model.
Two types of inflation is considered in the brane-world scenario
~\cite{low}.
The first type is bulk inflation.
If the moduli field $\phi$ in the bulk
has the appropriate potential $V(\phi)$,
it can cause the modular inflation~\cite{banks}.
The second type is brane inflation.
The matter fields in the brane cause the inflation.
In general, the inflation take place as a mixture of 
both types of the inflation~\cite{low}.

We obtained two types of the dynamical solutions ($\pm$) in $\S$IV.
It is interesting
to estimate which solution is most probable
by the framework of quantum cosmology.

There are the initial singularity at $\tau-\tau_0 =0$
and the singularity at $|y| \rightarrow \infty$
in the solutions which we have presented.
By solving the null geodesics equations,
we find that the $|y| =\infty$ singularity
occurs at the spatial infinity.
The 5-dimensional effective theory breaks down near the singularities.
It is thought that the initial singularity may be removed by quantum gravity
and quantum cosmology.
The singularity at $|y| \rightarrow \infty$ may be removed
in the model with the dimensions more than five 
($D>5$)~\cite{gib,hor}.
It is interesting issue to construct such a model.
From the view of AdS/CFT correspondence,
the curvature singularity at $|y| \rightarrow \infty$
may have physical meaning.
For example, in Ref.~\cite{gubser},
it is conjectured that 
large curvatures in geometries with the Minkowski brane
are physical
only if the scalar potential is bounded above
in the asymptotically AdS solution.
But the Gubser's conjecture can not be applied to our solutions
and is necessary to be generalized. 

While this work was being completed we became aware of related work by
Maeda and Wands,~\cite{mw} Mennim and Battye~\cite{mb} and
Barcelo and Visser~\cite{bv}.
We added the comment about this in $\S$IV and appendix.
%%%%%%%%%%%%%%%%%%%%%%%%%%%%%%%%%%%
\section*{Acknowledgments}
%%%%%%%%%%%%%%%%%%%%%%%%%%%%%%%%%%%
We are very grateful to Dr. Tetsuya Shiromizu for useful discussions.
This work was supported in part 
by the grant-in-Aid for Scientific Research (07CE2002) 
of the Ministry of Education, Science, Sports and Culture in Japan.

\appendix
%%%%%%%%%%%%%%%%%%%%%%%%%%%%%%%%%%%
\section*{}
%%%%%%%%%%%%%%%%%%%%%%%%%%%%%%%%%%%
We sketch the derivation of the effective equation on the brane
following Refs.~\cite{s,mw,mb,bv}
and give the useful expression of the $E_{\mu\nu}$ in the case with the radion $b$.
In the brane world scenario, the brane world is 
described by a thin domain wall ($(D-2)$-brane) $(M,q_{\mu\nu})$
in $D$-dimensional spacetime $(V,g_{\mu\nu})$. 
We denote the vector unit normal to $M$
by $n^\alpha$ and the induced metric on $M$ by 
$q_{\mu\nu} = g_{\mu\nu} - n_{\mu}n_{\nu}$.
We begin with the Gauss equation,
\begin{equation}
{}^{(D-1)}R^\alpha_{~\beta\gamma\delta}
= {}^{(D)}R^\mu_{~\nu\rho\sigma}
q^{~\alpha}_\mu q_\beta^{~\nu} q_\gamma^{~\rho}
 q_\delta^{~\sigma} + K^\alpha_{~\gamma}K_{\beta\delta} 
-K^\alpha_{~\delta}K_{\beta\gamma}\,.
\label{Gauss}
\end{equation}
where the extrinsic curvature  of $M$ is denoted by 
$K_{\mu\nu}= q_\mu^{~\alpha} q_\nu^{~\beta} \nabla_\alpha n_\beta$.
Contracting the Gauss equation~(\ref{Gauss}) on 
$\alpha$ and $\gamma$, we find
\begin{equation}
{}^{(D-1)}R_{\mu\nu}
= {}^{(D)}R_{\rho\sigma} q_\mu^{~\rho}q_\nu^{~\sigma} -  
{}^{(D)}R^\alpha_{~\beta\gamma\delta}n_\alpha q_\mu^{~\beta} n^\gamma 
q_\nu^{~\delta}  + KK_{\mu\nu} 
-K^{~\alpha}_{\mu}K_{\nu\alpha}\,,
\label{Ricci}
\end{equation}
where $K=K^\mu_\mu$ is the trace of the extrinsic curvature.
This readily gives
\begin{eqnarray}
{}^{(D-1)}G_{\mu\nu}
= \left[{}^{(D)}R_{\rho\sigma} -
{1 \over 2} g_{\rho\sigma}{}^{(D)}R\right] q_\mu^{~\rho} q_\nu^{~\sigma} 
+{}^{(D)}R_{\rho\sigma}n^\rho n^\sigma q_{\mu\nu}  
\\\nonumber
+ KK_{\mu\nu} 
-K^{~\rho}_{\mu}K_{\nu\rho} -{1 \over 2}q_{\mu\nu}  (K^2 
-K^{\alpha\beta}K_{\alpha\beta}) -{\tilde E}_{\mu\nu}\,, 
\label{4dEinstein-1}
\end{eqnarray}
where 
\begin{equation}
\tilde{E}_{\mu\nu} 
\equiv {}^{(D)}R^\alpha_{~\beta\rho\sigma}n_\alpha n^\rho 
q_\mu^{~\beta} q_\nu^{~\sigma} \,.
\end{equation}
Using the $D$-dimensional Einstein equations,
\begin{equation}
{}^{(D)}R_{\alpha\beta}
-\frac{1}{2}g_{\alpha\beta}{}^{(D)}R
= T_{\alpha\beta}\, ,
\label{5dEinstein}
\end{equation}
where  $T_{\mu\nu}$ is the $D$-dimensional energy-momentum tensor, 
together with the decomposition of the Riemann tensor into the Weyl
curvature, the Ricci tensor and the scalar curvature; 
\begin{eqnarray}
{}^{(D)}R_{\mu\alpha\nu\beta}
=\frac{D-3}{D-2}( g_{\mu [\nu}{}^{(D)}R_{\beta]\alpha}
-g_{\alpha [ \nu}{}^{(D)}R_{\beta] \mu})
\\\nonumber
-\frac{1}{2(D-3)}g_{\mu [\nu}g_{\beta ]\alpha}{}^{(D)}R
+{}^{(D)}C_{\mu\alpha\nu\beta}, 
\end{eqnarray}
we obtain the $(D-1)$-dimensional equations as
\begin{eqnarray}
{}^{(D-1)}G_{\mu\nu}
= {D-3 \over D-2}\left(T_{\rho\sigma}
q^{~\rho}_{\mu}  q^{~\sigma}_{\nu}  
+\left(T_{\rho\sigma}n^\rho n^\sigma-{1 \over D-1}T^\rho_{~\rho}\right)
 q_{\mu\nu} \right) 
\nonumber \\
+ KK_{\mu\nu} 
-K^{~\sigma}_{\mu}K_{\nu\sigma} -{1 \over 2}q_{\mu\nu}
  \left(K^2-K^{\alpha\beta}K_{\alpha\beta}\right) - E_{\mu\nu}, 
\label{4dEinstein}
\end{eqnarray}
where
\begin{equation}
E_{\mu\nu} \equiv {}^{(D)}C^\alpha_{~\beta\rho\sigma}n_\alpha n^\rho 
q_\mu^{~\beta} q_\nu^{~\sigma} .
\label{Edef}
\end{equation}
Note that $E_{\mu\nu}$ is traceless.

The dilaton field equation in the bulk is reduced to
\begin{equation}
\Box\phi-n^{\nu}(\nabla_{\nu}n^{\mu})\nabla_{\mu}\phi
+K\mbox \pounds_n \phi +\mbox \pounds_n^2\phi
-\frac{\lambda\Lambda}{8}e^{\lambda\phi/(D-2)}=0.
\end{equation}

The junction condition for the extrinsic curvature and
$Z_2$ symmetry tells us
that
the extrinsic curvature on the brane is uniquely written
in the terms of the energy-momentum tensor $S_{\mu\nu}$
on the brane~\cite{is};
\begin{equation}
K_{\mu\nu}|_{brane}=-\frac{1}{2}(S_{\mu\nu}-\frac{1}{D-2}q_{\mu\nu}S).
\label{k}
\end{equation}
In our system the energy-momentum tensor $S_{\mu\nu}$ on the brane
is given by
\begin{equation}
S_{\mu\nu}=-\frac{V}{2}e^{\lambda\phi/2(D-2)}q_{\mu\nu}.
\end{equation}
%We obtain 
%\begin{equation}
%K_{\mu\nu}=-\frac{1}{4(D-2)}Ve^{\lambda\phi/2(D-2)}q_{\mu\nu}.
%\end{equation}
The junction condition for the dilatonic field and $Z_2$ symmetry imply us
that the derivative of the dilaton field
with the coordinate $y$ of the fifth dimension on the brane is written as
\begin{equation}
\phi'|_{brane}=\frac{D-2}{16}Ve^{\lambda\phi/(D-2)},
\label{phi}
\end{equation}
where $\phi'|_{brane}=\lim_{\epsilon\rightarrow +0}\phi'(\epsilon)$.

Substituting ~(\ref{k}) into ~(\ref{4dEinstein}) and
using ~(\ref{phi}), we obtain
the $(D-1)$-dimensional effective equations on the brane 
(Eq.~(\ref{4dEinstein2})).
The effective equation for the dilaton field is given by
\begin{equation}
\Box\phi=\frac{4(D-2)\Lambda+V^2}{32(D-2)}\lambda e^{\lambda\phi/(D-2)}
-\phi''|_{brane}.
\end{equation}

We have assumed the $D$-dimensional metric to have the form,
\begin{equation}
ds^2=b^2 dy^2+q_{\mu\nu}dx^\mu dx^\nu\,.
\end{equation}
It is useful to write down the following formula
for $E_{\mu\nu}$:
\begin{equation}
E_{\mu\nu}  = 
 {\tilde E}_{\mu\nu}-\frac{1}{D-2}q_{\mu\nu}{}^{(D)}R_{\alpha\beta}
n^\alpha n^\beta
-\frac{1}{D-2}q_\mu^\alpha q_\nu^\beta {}^{(D)}R_{\alpha\beta}
+\frac{1}{(D-1)(D-2)}q_{\mu\nu}{}^{(D)}R,
\end{equation}
where
\begin{eqnarray}
{\tilde E}_{\mu\nu}&\equiv&{}^{(D)}R_{\mu \alpha\nu\beta}n^\alpha n^\beta
\nonumber \\
&=&-\mbox \pounds_n K_{\mu\nu}+K_{\mu\alpha}K_\nu^\alpha\
-D_{\nu}D_{\mu}\log b 
-D_{\mu}\log b D_{\nu}\log b,
\end{eqnarray}
and $D_\mu$ is 
the covariant differentiation with respect to $q_{\mu\nu}$.
%%%%%%%%%%%%%%%%%%%%

\end{document}